\documentclass{article}

\usepackage{PRIMEarxiv}

\usepackage[utf8]{inputenc} % allow utf-8 input
\usepackage[T1]{fontenc}    % use 8-bit T1 fonts
\usepackage{hyperref}       % hyperlinks
\usepackage{url}            % simple URL typesetting
\usepackage{booktabs}       % professional-quality tables
\usepackage{amsfonts}       % blackboard math symbols
\usepackage{nicefrac}       % compact symbols for 1/2, etc.
\usepackage{microtype}      % microtypography
\usepackage{lipsum}
\usepackage{fancyhdr}       % header
\usepackage{graphicx}       % graphics
\graphicspath{{media/}}     % organize your images and other figures under media/ folder
\usepackage{subcaption}
\usepackage{tabularx}
\usepackage{amsmath}
\title{ Energy Prediction of PV Panels for Demand and Response System Using ANN (Deep Learning)
}

\author{
Rohaib Bhatti \\
    National University Of Sciences and Technology\\
    Islamabad\\
   \texttt{rbhatti.ee41ceme@student.nust.edu.pk} \\
   \And
  Ali John Naqvi \\
  National University of Sciences and Technology \\
  Islamabad\\
  \texttt{salijohn.ce41ceme@student.nust.edu.pk} \\
   \And
  Abdullah Tauqeer \\
  National University Of Sciences and Technology \\
  Islamabad\\
  \texttt{atauqeer.ee41ceme@student.nust.edu.pk} \\
}

\begin{document}
\maketitle

\begin{abstract}
{Renewable sources of energy are the future due to the environmental problems caused by non-renewable sources to produce energy. The biggest issue with renewable energy sources is that the power produced by devices such as PV solar panels depend on many uncertain factors. These factors include Solar irradiation, wind speed, temperature, hours of sunlight per day, and surface temperature of solar panels. Industries and authorities can use this predicted power through ML to control power consumption. Power forecast has multiple applications that promote the usage of green energy in the future. This paper will also help to determine the dependence of PV power production on various weather/environmental factors. For this paper, we have used regressions and ANN models to predict power. In the end, results of power prediction using regression as well as the ANN model are compared with the actual power output. Overall, ANN performs excellently compared to the other machine learning models because of its advanced feature selection techniques.}
\end{abstract}

% keywords can be removed

\section{Introduction}

The motivation to write this research comes from the aim of contributing towards a sustainable world. A sustainable world is not possible without producing green energy. This research aims to solve the problem of uncertainty of the power being produced through solar PV panels. Providing accurate predictions of solar power generation will help industries and power authorities maintain an uninterruptible power supply by demand and response system. This study will result in more power being produced by renewable sources than non-renewable, thus saving the environment from global warming and climate change.

In the future, the use of solar energy will grow in enormous numbers because it is environmentally friendly, especially in regions like Pakistan, where there is great potential for producing PV power. In Pakistan, most areas have extended hours of sunlight availability throughout the year, especially in the country's southern region [3]. It is not wrong to say that solar energy is the most abundant energy source that is available to our planet, and it is something that will not vanish soon. 100,000 TW of electrical energy can be produced through sunlight every hour, which is such a huge amount that it can give power to the entire population of humans on earth for a year [2]. This statistic shows that humans should take advantage of that and produce green energy that can meet our requirements and not damage the earth's environment.

Solar panels are easy to install because no complex setup or massive construction is needed. Overall, these cells are also easy to manufacture for industries, which shows that there is great potential for developing nations like Pakistan to fulfill their energy needs using an economical energy source such as sunlight [5]. The most significant disadvantage of using solar power is that it requires many areas to install PV panels to produce a megawatt of electricity. Currently, Pakistan's energy needs are 92 GWs, and Pakistan has the potential to produce 2.9 TW of solar electricity, which shows that solar alone can meet all of the country's energy requirements, and still, much energy will be left to export [2].

The system is also clearly suited to off-grid generating and consequently to places with little infrastructure. The public accepts and generally approves solar technology; it is subject to less geopolitical, environmental, and aesthetic concerns than nuclear, wind, or hydro. However, substantial desert installations may arouse objections [6]. 

Photovoltaics are challenging to integrate into power systems since solar energy is mainly reliant on location and weather, shifting wildly. This uncertainty results in system disturbances, voltage spikes, configuration issues, ineffective utility management, and economic loss. Forecast models can be helpful, but time stamps, forecast horizons, input correlation analysis, data pre-and post-processing, weather categorization, capacity planning, risk assessment, and performance assessments must all be taken into account [1].

One of the significant issues the industry is presently facing is incorporating these unique types of power plants into the electrical system. Because renewable energy is becoming a more significant part of the power market, machine learning algorithms must forecast foreseeable electricity production to a desired level of accuracy. These projections also apply to power plant owners, the energy trading market, and grid operators. Information on future energy production growth in the grid decreases all market players' technical and economic risks [4].

The fundamental objective of this work is to evaluate the use of ML and deep learning techniques to predict solar irradiance. The paper will also go through the intricacies of the forecasting methodologies with illustrations. Each approach's merits and flaws are also discussed. The predicted data and actual data comparisons based on our simulations are explored. Furthermore, intriguing future research paths are suggested.

\section{Method}
The methodology used in this research will be using an ML model, a regression model, and a deep learning model, ANN. Using the data from the world bank about environmental factors of Pakistan, we trained a regression model, as well as training of the artificial neural network was also done. Four solar panels were installed in Islamabad to take real-time data and compare it with forecasted data to get an idea of the accuracy of our predicted data.

\vspace{0.4cm}
\begin{figure}[h]
 \centering
  \includegraphics[width= 10cm, height= 6cm]{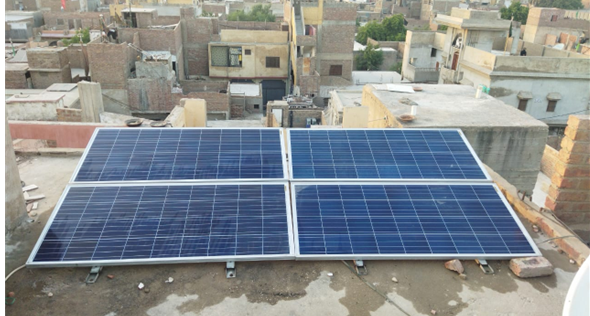}
  \caption{PV Setup used to predict power}
  \label{fig:Setup}
\end{figure}

These four plates were installed in Islamabad, and the world bank's data was also from Islamabad. Four of these PV solar panels were each 150w. The nameplate of the solar cell is attached in figure 2.

\begin{figure}[h]
 \centering
  \includegraphics[width= 10cm, height= 11cm]{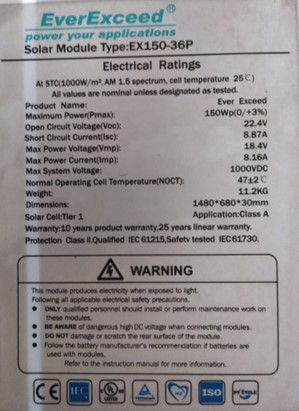}
  \caption{Nameplate of PV panel}
  \label{fig:Name plate}
\end{figure}

The 4 PV Panels of solar are of EverExceed, and their module type is EX150-36P. The maximum power Pmax they can generate is 150W. These panels can operate at temperatures 47 $\pm$ 2. The dimensions of the solar panels are 1480*680*30mm. The max power voltage it can produce is 18.4V, and the max power current that can be produced is 8.16 A.

\subsection{Data}
The world bank data set for weather conditions was acquired as they had weather data for the capital of Pakistan, "ISLAMABAD," for 4 years, starting from 2014 to 2017, with a resolution of 10 minutes. Data was factorized into 11 features which included the following: 

\begin{table}[h]
\centering
\caption{Features of Data set. (WorldBank, 2021)}
\vspace{0.1cm}
\begin{tabular}{|c|c|}
\hline
Time & Date and time according to ISO8601 (yyyy-mm-dd HH: MM)
\\
 \hline
 ghi pyr	& Global horizontal irradiance (W/m²) from thermopile pyranometer\\
 \hline
 Dni & Direct normal irradiance (W/m²) from thermopile pyrheliometer\\
 \hline
 air temperature	& Air temperature (°C) at 2 m height\\
 \hline
relative humidity &	Relative humidity (\%) at 2 m height\\
 \hline
 wind speed	& Wind speed (m/s) at 10 m height \\
 \hline
wind speed of gust&	Maximum wind speed in the integration interval \\
 \hline
 wind from direction st dev &	The wind direction in degrees north counted clockwise (standard deviation)\\
 \hline
 wind from direction	& The wind direction in degrees north counted clockwise\\
 \hline
 barometric pressure	& Ambient air pressure in hPa\\
 \hline
 sensor cleaning &	1 (yes) / 0 (no)\\
 \hline
\end{tabular}
\end{table}

Using the features given above, following parameters were chosen to predict power from these 4 solar PV panels. 

\begin{table}[h]
\centering
\caption{Input Parameters of the Model}
\vspace{0.1cm}
\begin{tabular}{|c|c|c|c|}
\hline
Environmental Parameters  & Min & Max & Unit \\
 \hline
 Air temperature & 0.8 & 43.3 & C\\
 \hline
 Wind speed & 0 & 17.6 & m/s\\
 \hline
 Wind direction & 0 & 360 & Degrees\\
 \hline
 Power Output & 0 & 600 & W\\
 \hline
 Irradiance & 0 & 1140 & $\frac{W}{m^2}$ \\
 \hline
 Relative humidity & 9.1 & 100 &\% \\
 \hline
\end{tabular}
\end{table}

\section{Research Procedure}
%section text
Firstly, 2014 to 2016 is used to train regression and ANN models. After that, data testing is carried out on the data from 2017. Testing data was kept separate from the training to ensure that testing of the model was done correctly. The GPU used to do the training of models was Google Collab. Both ANN and regression models were trained on Google Collab

The formula used to calculate power from the forecasted factors is:

\begin{equation}
    E = A \times r \times H \times PR
\end{equation}

Where E is energy, A is Total Area of the panel, r is the yield of the solar panel, H is the global solar irradiation, and PR is the performance ratio. Using this formula, predicted power is calculated. 

After model training and testing, current and voltage sensors were installed at the output of the combined power of the four PV panels. The output of the actual power production was acquired for the whole day of June 12, 2022, and it was compared with the power predicted for June 12, 2022. June was chosen as the month of the forecast because it is the hottest month in Pakistan providing intense sunlight for long hours throughout the day, helping to generate PV power. Secondly, the output of actual power production was again taken for December 12, 2022, and compared with the predicted power output for December 12, 2022. 

To calculate the accuracy of the model, we have employed the absolute percentage formula, which is:

\begin{equation}
    APE = \frac{|V_f - V_a|}{V_a} \times 100\%
\end{equation}

In this formula, Vf is the forecasted power that we predicted from ANN and regression models. Va is the actual power produced by the 4 solar panels.
The results of actual power and predicted power using ANN, as well as regression, are shown below:

\begin{figure}[h]
  \centering
  \begin{subfigure}[b]{0.44\linewidth}
    \includegraphics[width=7cm, height= 5cm]{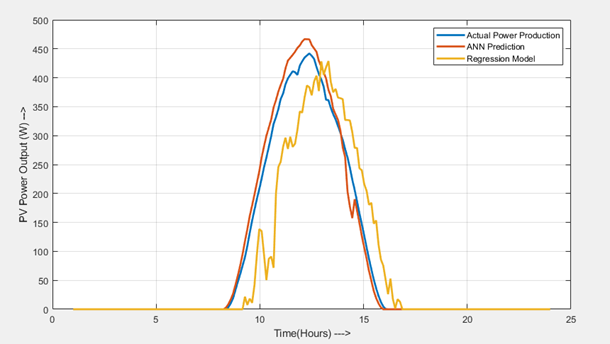}
    \caption{}
  \end{subfigure}
  \begin{subfigure}[b]{0.44\linewidth}
    \includegraphics[width=7cm, height=5cm]{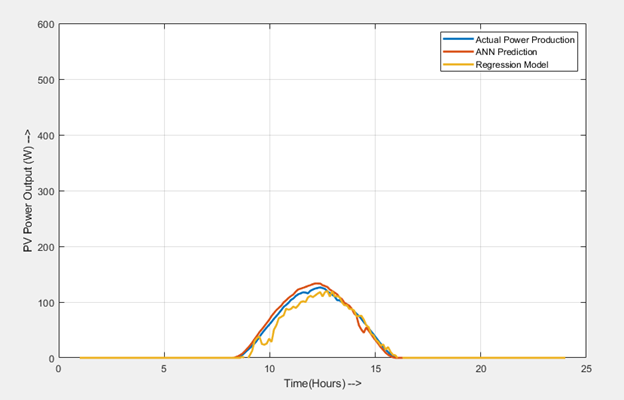}
    \caption{}
  \end{subfigure}
  \caption{(a)Actual and predicted PV power in summer. (b) Actual and predicted PV power in winter}
  \label{fig:comparison}
\end{figure}

Figures 3 shows that ANN has predicted quite close to the actual power produced from the PV panels. Although the regression model has a more significant percentage error, it still follows the trend quite well. The credibility of the prediction is further verified by the difference in power prediction between June and December. Since there is less solar irradiance and intense hours of sunlight in December, the energy prediction has declined the same as the actual power production.

\begin{figure}[h]
 \centering
  \includegraphics[width= 12cm, height= 7cm]{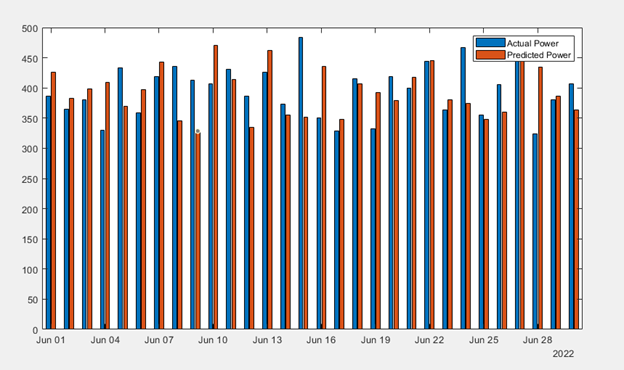}
  \caption{Comparison between prediction and actual power for June 2022}
  \label{fig:Comparison}
\end{figure}

Figure 4 compares the actual and the predicted power for the entire month of June. It shows that the predicted power has followed the same trend as the actual power throughout the month. The accuracy levels have fluctuated, but the machine learning model will learn and improve based on feedback. The percentage error for each prediction is also shown in figure 5:

\begin{figure}[h]
 \centering
  \includegraphics[width= 12cm, height= 7cm]{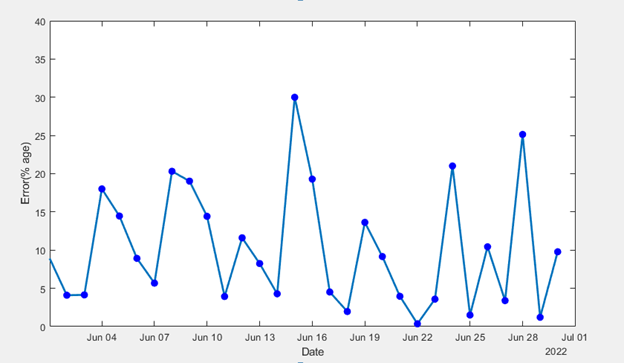}
  \caption{Absolute percentage error for each day of June 2022}
  \label{fig:error}
\end{figure}

The most significant percentage error can be seen on June 15, when ANN predicted almost 30 percentage less power than the actual power output.

\section{Conclusion}
In conclusion, we can state that ANN is a far superior model when it comes to the prediction of PV power as compared to the regression model. Power prediction overall has been entirely satisfactory and near the actual power, which is a significant achievement for this research. The prediction also has vast applications for demand and response systems. With the power generation for the next day already known, power consumption can be managed accordingly to avoid a blackout. This prediction is also helpful for investors in solar energy stations as they can precisely predict the future electricity generation and make decisions of break even and profitability based on that up to some level of credibility.
My suggestion for future work is to use a combination of ML models to predict PV power in Pakistan because the average of different models used might provide better accuracy than a single model. Increased accuracy is essential for future renewable energy in Pakistan and worldwide. Another suggestion is that the application of demand response should be made using forecasts to adjust loads accordingly and avoid blackouts.

%Bibliography
\bibliographystyle{unsrt}  
\bibliography{references}  
[1]Ahmed, R., Sreeram, V., Mishra, Y.,  Arif, M. D. (2020). A review and evaluation of the state-of-the-art in PV solar power forecasting: Techniques and optimization. Renewable and Sustainable Energy Reviews, 124, 109792. https://doi.org/10.1016/j.rser.2020.109792

[2]Ali, S., Yan, Q., Sajjad Hussain, M., Irfan, M., Ahmad, M., Razzaq, A., Dagar, V.,  Işık, C. (2021). Evaluating Green Technology Strategies for the Sustainable Development of Solar Power Projects: Evidence from Pakistan. Sustainability, 13(23), 12997. https://doi.org/10.3390/su132312997

[3]Bacher, P., Madsen, H.,  Nielsen, H. A. (2009). Online short-term solar power forecasting. Solar Energy, 83(10), 1772–1783. https://doi.org/10.1016/j.solener.2009.05.016

[4]Gensler, A., Henze, J., Sick, B.,  Raabe, N. (2016, October 1). Deep Learning for solar power forecasting — An approach using AutoEncoder and LSTM Neural Networks. IEEE Xplore. https://doi.org/10.1109/SMC.2016.7844673

[5]Ren, Y., Suganthan, P. N.,  Srikanth, N. (2015). Ensemble methods for wind and solar power forecasting—A state-of-the-art review. Renewable and Sustainable Energy Reviews, 50, 82–91. https://doi.org/10.1016/j.rser.2015.04.081

[6]Wan, C., Zhao, J., Song, Y., Xu, Z., Lin, J., and Hu, Z. (2015a). Photovoltaic and solar power forecasting for smart grid energy management. CSEE Journal of Power and Energy Systems, 1(4), 38–46. https://doi.org/10.17775/cseejpes.2015.00046

[7]WorldBank. (2021). Data Catalog. Datacatalog.worldbank.org. https://datacatalog.worldbank.org/search/dataset/0038550

\end{document}